\documentclass[aps,epjc,twocolumn,floatfix,superscriptaddress,nofootinbib,showpacs,showkeys]{revtex4}
\usepackage{graphicx}
\usepackage{dcolumn}  % Align table columns on decimal point
\usepackage{bm}       % bold math
\usepackage{palatino}
\begin{document}
\input{epsf}

\title{Nuclear attenuation of high energy multi-hadron systems  in the string model}
%\date{today}
\author{L.~Grigoryan}
\affiliation{Yerevan Physics Institute, Br.Alikhanian 2, 375036 Yerevan, Armenia}
\email{leva@mail.desy.de}

%%%%%%%%%%%%%%%  begin abstract  %%%%%%%%%%%%%%%%%%%%%%%%%%%%%%%
\begin{abstract}
\hspace*{1em} Nuclear attenuation of the multi-hadron systems in the string model
is considered. The improved two-scale model with set of parameters
obtained recently for the single hadron attenuation is used for calculation of
the multiplicity ratios of the one-, two- and three-hadron systems electroproduced
on nuclear and deuterium targets. The comparison of the features of the one-, two-
and three-hadron systems is performed. The predictions of the model for multiplicity 
ratios of multi-hadron systems as functions of different convenient variables
are presented.
\end{abstract}
\maketitle
%%%%%%%%%%%%%%%  end abstract  %%%%%%%%%%%%%%%%%%%%%%%%%%%%%%%
%\twocolumn  
%\bf   
%\Large
%\vspace{-20.0cm}
%%%%%%%%%%%%%%%  begin \section{Introduction}  %%%%%%%%%%%%%%%%%%%%%%%%%
\section{Introduction}
\normalsize 
\hspace*{1em}
In any hard process the initial interaction takes place between partons,
which then turn into the final hadrons by means of hadronization process.
However the hadronization process cannot be described in the framework of the 
existing theory of the strong interactions (perturbative QCD), because of major
role of "soft" interactions. Therefore the experimental and theoretical (on the
level of phenomenological models) studies of the all aspects of the transition
from partons to hadrons are very important. The space-time evolution of the 
hadronization process, despite its importance, has been studied relatively little. 
The study of the early stage of the hadronization process can shed additional
light on the further development of the process. Semi-inclusive reactions with 
nuclear targets give the possibility to study the development of the hadronization 
process on distances of a few Fermi from the point of initial interaction.\\
\hspace*{1em}
In particular, the nuclear attenuation (NA) of the high energy hadrons is the
well known tool for investigation a early stage of hadronization process
\footnote{NA means the difference of the ratio the multiplicities (per nucleon)
on nucleus to that on deuterium from unity.}.
There are many phenomenological models, which describe, rather qualitatively,  
existing experimental data for single hadron NA~\cite{A1}-\cite{A11}. Also 
some predictions for the attenuation of multi-hadron systems leptoproduced in 
nuclear matter in the framework of the string model were done~\cite{A12}-\cite{A13}.
It was argued that measurements of NA of a multi-hadron systems can remove 
some ambiguities in determination of the parameters describing strongly 
interacting systems at the early stage of particle production: formation time 
of hadrons and cross-section for the intermediate state to interact inside the 
nucleus. Then, for the first time, data on the two-hadron system multiplicity
ratio were obtained in electroproduction~\cite{A14}. Experiment was performed
in specific conditions. The multiplicity ratio of the charged hadrons was
measured as a function of the fractional energy of the subleading hadron $z_{2}$,
whereas over the fractional energy of the leading hadron $z_{1}$ the integration
in the region $0.5 < z_{1} < 1 - z_{2}$ was performed. Later the data on the 
two-hadron system multiplicity ratio in neutrinoproduction were presented by
another experiment~\cite{A15}.\\
\hspace*{1em}
The data on the two-hadron system multiplicity ratio~\cite{A14} were described in
the framework of some theoretical models: the probabilistic coupled-channel transport
model~\cite{A16}, the so called energy loss model~\cite{A17} and the string
model~\cite{A18}. In particular we showed~\cite{A15,A18}, that based on the two-scale
model (TSM)~\cite{A4} and improved two-scale model (ITSM)~\cite{A10}, it is possible
to describe these data quantitatively in the framework of the string model.
We presented also predictions for the dependence of two-hadron system NA on the
virtual photon's energy in the same model.
%%% old text %%%
%Possible mutual screening of the
%hadrons in string and its experimental verification were discussed
%\footnote{The term "string" here means the object arising in result of DIS, which
%during its space-time evolution turns into states consisting of strings, prehadrons
%and hadrons. After all this object turns into the jet of hadrons.}.\\
%%% new text %%%
Possible mutual screening of the hadrons occurred from one string and its 
experimental verification have been discussed.\\
\hspace*{1em}
In this work we continue the study the electroproduction of multi-hadron systems in
cold nuclear matter. This is the main goal of the present paper to consider
the mutual screening of the prehadrons and hadrons in string (jet)
\footnote{The term "string" here means the object arising in result of DIS, which
during its space-time evolution turns into states consisting of strings, prehadrons
and hadrons. After all this object turns into the jet of hadrons.}.
We compare one-, two- and three-hadron systems and show
that mutual screening of prehadrons and hadrons plays essential role and
can be measured experimentally. For instance such data can be obtained
by HERMES Experiment, SKAT Experiment, and JLab after upgrade
to the energy 12 GeV. We suppose that investigation of the mutual screening of 
prehadrons and hadrons in cold nuclear matter can help to establish initial
conditions for the study of similar processes in hot nuclear matter arising
in high energy hadron-nucleus and nucleus-nucleus interactions at RHIC and LHC.\\
\hspace*{1em}
The paper is organized as follows. In Section II the theoretical framework is
briefly described. Results and discussion as well as necessary ingredients for 
calculations are presented in Section III. Our conclusions are given in Section IV.
%%%%%%%%%%%%%%%  end \section{Introduction}  %%%%%%%%%%%%%%%%%%%%%%%%%
%\bf
%\Large
%%%%%%%%%%%%%%%  begin \section{Theoretical framework}  %%%%%%%%%%%%%%%
\section{Theoretical framework}
\normalsize
\hspace*{1em}
In the works~\cite{A12}-\cite{A13} the process of leptoproduction of multi-hadron
systems on a nucleus with atomic mass number $A$ was considered theoretically for
the first time. Although in these papers discussions were presented for the
general case of the $n$ hadrons observed in the final state, and some formulae
have been written for this general case, however the basic formulae and numerical
calculations were performed for the case of two observed hadrons. In this work we
make next step in this direction and consider three hadron systems observed in
the final state of the high energy semi-inclusive lepton-nucleus interaction.
 %%%%%%%%%%%%%%%%%%%%%%%%  begin thr_hadfig1.eps  %%%%%%%%%%%%%%%%%%%%%%%%%%%%
%\begin{figure}[!htb]
\begin{figure}[!t]
\begin{center}
\epsfxsize=8.cm
\epsfbox{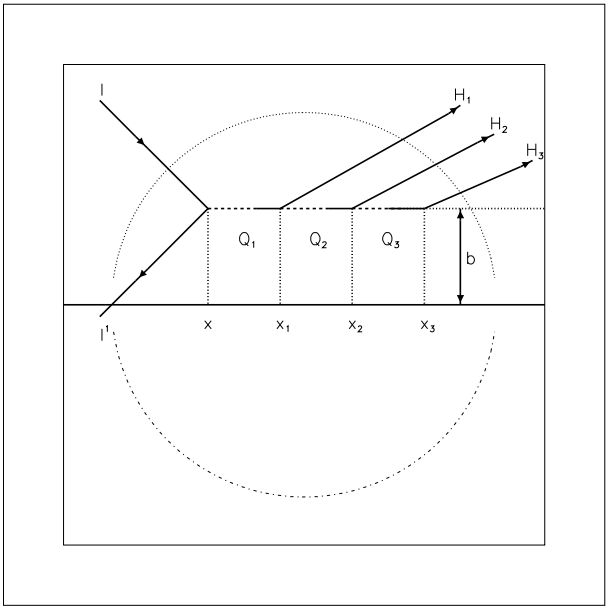}
\end{center}
\caption{\label{xx1}
{\it Leptoproduction of three-hadron system on nuclear target. Details see in
 text.}}
\end{figure}
%%%%%%%%%%%%%%%%%%%%%%%%  end thr_hadfig1.eps  %%%%%%%%%%%%%%%%%%%%%%%%%%%%
We do not give
the equations for the case of one or two hadrons, which will also be used for 
calculations and discussions in the article because the conversion of these
formulae to the case of one or two hadrons are very simple. The semi-inclusive
reaction of the leptoproduction of three hadrons on nuclear target is:
\begin{eqnarray}
          {l_i + A \rightarrow l_f + h_1 + h_2 + h_3 + X\hspace{0.15cm},}
\end{eqnarray}
where $l_i (l_f)$ is the initial (final) lepton, $h_1$, $h_2$ and $h_3$ are the
observed hadrons.
The hadrons $h_1$, $h_2$ and $h_3$ carry fractions $z_1$, $z_2$ and
$z_3$ of the total available energy (the energy conservation implies the condition:
$z_1+z_2+z_3 \le 1$).
The multiplicity ratio for that
process is defined as (it is assumed that averagings over transverse momenta of the 
final hadrons performed).
\begin{eqnarray}
\nonumber
{R_M^{3h} = 2d\sigma_A(\nu,Q^2,z_1,z_2,z_3)/}
\end{eqnarray}
\begin{eqnarray}
{Ad\sigma_D(\nu,Q^2,z_1,z_2,z_3)\hspace{0.15cm},} 
\end{eqnarray}
where $d\sigma_A$ and $d\sigma_D$ are the cross-sections for the reaction (1)
on nuclear and deuterium targets, respectively, $\nu$ denotes the energy of the 
virtual photon and $Q^{2}=-q^{2}$, where $q^{2}$ is the square of the four-momentum 
of the virtual photon. One can imagine the reaction (1) as shown in Fig.1. The
interaction of the lepton with the intranuclear nucleon occurs at the point $(b,x)$ 
from which the intermediate state $q$ begins its propagation ($b$ and $x$ are the
impact parameter and the longitudinal coordinate of the DIS point). Initially the
intermediate state q presents itself the object like a string with knocked-out
quark on the fast and nucleon remnant on slow ends, which connected by means of
string consisting of gluons. During further movement string breaks on the smaller
pieces, and in result at the points $(b,x_{c1})$, $(b,x_{c2})$ and $(b,x_{c3})$ the 
first constituents (valence quarks or antiquarks) of the hadrons $h_1$, $h_2$ and
$h_3$ are produced, and at the points $(b,x_{1})$, $(b,x_{2})$ and $(b,x_{3})$ the 
second constituents are produced and the yo-yo of the hadrons $h_1$, $h_2$ and $h_3$
arise (the term "yo-yo" means, that the colorless system with valence contents and 
quantum numbers of the final hadron is formed, but without its "sea" partons).
The points $(b,x_{c1}), (b,x_{c2})$ and $(b,x_{c3})$ do not represented in figure,
but they are used properly in the calculations. In Fig.1 we for the sake of
simplicity represent the case of three adjacent hadrons. In fact, all possibilities
have been considered in the calculations both adjacent and not adjacent hadrons.\\
\hspace*{1em}
We do not take into account the hadrons produced in result of decay the
resonances. This factor could lead to an increase of nuclear attenuation, if taken 
into account only one hadron from each resonance and a decrease of nuclear
attenuation, if taken into account that two or three hadrons can be produced from
the same resonance. We think that the overall effect is small.\\
\hspace*{1em}
In the string model there are simple connections between above mentioned points
$x_{1}-x_{c1}=z_1L$, $x_{2}-x_{c2}=z_2L$ and $x_{3}-x_{c3}=z_3L$, where 
$L$ is the full hadronization length, $L = \nu/\kappa$,\hspace{0.15cm}$\kappa$
is the string tension ($\kappa=1 GeV/fm$). The multiplicity ratio for the case of
three hadrons observed in the final state can be presented in the form:
\begin{widetext}
\begin{eqnarray}
\nonumber
{R_M^{3h}\approx\frac{1}{6} \int{d^2b} \int_{-\infty}^{\infty}dx
\int_{x}^{\infty}dx_1\int_{x_1}^{\infty}dx_2\int_{x_2}^{\infty}
{dx_3\rho(b,x)\times}}
\end{eqnarray}
\begin{eqnarray}
\nonumber
{[D(z_1,z_2,z_3,x_1-x,x_2-x,x_3-x)W_0(h_1,h_2,h_3;b,x,x_1,x_2,x_3)+}
\end{eqnarray}
\begin{eqnarray}
\nonumber
{D(z_1,z_3,z_2,x_1-x,x_2-x,x_3-x)W_0(h_1,h_3,h_2;b,x,x_1,x_2,x_3)+}
\end{eqnarray}
\begin{eqnarray}
\nonumber
{D(z_2,z_1,z_3,x_1-x,x_2-x,x_3-x)W_0(h_2,h_1,h_3;b,x,x_1,x_2,x_3)+}
\end{eqnarray}
\begin{eqnarray}
\nonumber
{D(z_2,z_3,z_1,x_1-x,x_2-x,x_3-x)W_0(h_2,h_3,h_1;b,x,x_1,x_2,x_3)+}
\end{eqnarray}
\begin{eqnarray}
\nonumber
{D(z_3,z_1,z_2,x_1-x,x_2-x,x_3-x)W_0(h_3,h_1,h_2;b,x,x_1,x_2,x_3)+}
\end{eqnarray}
\begin{eqnarray}
%\nonumber
{D(z_3,z_2,z_1,x_1-x,x_2-x,x_3-x)W_0(h_3,h_2,h_1;b,x,x_1,x_2,x_3)],} 
\end{eqnarray}
where $D(z_1,z_2,z_3,l_1,l_2,l_3)$ (with $l_1 < l_2 < l_3$) is the distribution
of the constituent formation lengths $l_1$, $l_2$ and $l_3$ of the hadrons and
$\rho(b,x)$
is the nuclear density function normalized to unity. $W_0$ is the probability
that neither the hadrons $h_1$, $h_2$, $h_3$ nor intermediate states leading to
their production (initial strings) interact inelastically in nuclear matter: 
\begin{eqnarray}
\nonumber
{W_0(h_1,h_2,h_3;b,x,x_1,x_2,x_3) =(1-Q_1-(H_1+Q_2+H_2+Q_3+H_3-}
\end{eqnarray}
\begin{eqnarray}
%\nonumber
{H_1(Q_2+H_2+Q_3+H_3-H_2(Q_3+H_3))-H_2(Q_3+H_3)))^{(A-1)}\hspace{0.15cm},}       
\end{eqnarray}
where $Q_1$, $Q_2$ and $Q_3$ are the probabilities for the initial strings of
the corresponding hadrons to be absorbed in the nucleus within the intervals
$(x,x_1)$, $(x_1,x_2)$ and $(x_2,x_3)$, respectively. $H_i\hspace{0.15cm}(i=1,2,3)$
is the probability for the $h_i$ to interact inelastically in nuclear matter,
starting from point $x_{i}$. The probabilities $Q_1, Q_2, Q_3, H_1, H_2, H_3$ can
be calculated using the general formulae:
\begin{eqnarray}
 {P(x_{min},x_{max})=\int_{x_{min}}^{x_{max}}\sigma_P\rho(b,x)dx\hspace{0.15cm},}  
\end{eqnarray}
where the subscript $P$ denotes the particle (initial string or hadron), $\sigma_P$
its inelastic cross section on nucleon target, and $x_{min}$ and $x_{max}$ are the
end points of its path in the x direction, as it is shown in Fig.1.\\
\hspace*{1em}
We use the scaling function of the standard Lund model for calculations. The simple
form of this function $f(z) = (1 + c)(1 - z)^{c}$, where $c \approx 0.3$ is the 
parameter which controls the steepness of the standard Lund model's fragmentation
function, allows to sum the sequence of produced hadrons over all ranks and to obtain 
the analytic expression for the any number of particles observed in final state.
In the general case of the $n$ hadrons the distribution $D(z_1,...,z_n;l_1,...,l_n)$ 
of the constituent formation lengths $l_1$,...,$l_n$ is:
\begin{eqnarray}
\nonumber
{D(z_{1} \cdot\cdot\cdot z_{n};l_{1} \cdot\cdot\cdot l_{n})=L^{n}(1+c)^{n}\frac{(l_{1} 
\cdot\cdot\cdot l_{n})^{c}}
{((l_{1}+z_{1}L) \cdot\cdot\cdot (l_{n}+z_{n}L))^{1+c}}\bigg[\delta(l_{n}-(1-z_{n})L) 
+\frac{1+c}{l_{n}+z_{n}L}\bigg]}
\end{eqnarray}
\begin{eqnarray}
%\nonumber
{\times \bigg[\delta(l_{n}-l_{n-1}-z_{n-1}L)+\frac{1+c}{l_{n-1}+z_{n-1}L}\bigg]
\cdot\cdot\cdot \bigg[\delta(l_{2}-l_{1}-z_{1}L)+\frac{1+c}{l_{1}+z_{1}L}\bigg],}
\end{eqnarray}
where $l_{n} \le (1-z_{n})L$, $l_{n-1} \le l_{n}-z_{n-1}L$,...,
$0 \le l_{1} \le l_{2}-z_{1}L$.
Equation (6) was obtained for the first time in Ref.~\cite{A13}. Unfortunately,
corresponding equation (2.21) from Ref.~\cite{A13} contains some mistakes and
uncertainties.
\end{widetext}
%%%%%%%%%%%%%%%  end Theoretical framework  %%%%%%%%%%%%%%%%%%%

%%%%%%%%%%%%%%%  begin \section{Results and Discussion}  %%%%%%%%%%%
\section{Results and Discussion}

\hspace*{1em}
Multi-hadron production depends on many variables. This complicates the study
of such systems. For example in the electroproduction process when in final state
are observed $n$ hadrons, even after averaging over virtuality of photon and
transverse momenta of hadrons the ratio of multiplicities depends on $n+1$ variables
(the fractional energies of hadrons and energy of photon). Although we restrict
ourself in this paper by three hadrons observed in final state, the simultaneous
consideration of four variables is very difficult especially in experimental study.
We escape this difficulty by inclusion of some additional averagings. We will
consider following combinations of fixed and averaged variables: (i) the dependence
on the fractional energy of one of the hadrons, the "trigger" hadron ($z_{tr}$),
whereas integrations are performed over the fractional energies of other hadrons
and the energy of virtual photon $\nu$ is kept at fixed value (in this paper it is
fixed at value $10 GeV$); (ii) the dependence on the number of observed hadrons
$n$, whereas the
averagings are performed over $z_{tr}$ in some regions and $\nu$ is kept fixed;
(iii) the $\nu$-dependence at fixed value of the "trigger" hadron fractional energy
$z_{tr}=0.3$; (iv) the dependence on the fractional energy of multi-hadron system
$Z=\sum_{i=1}^{n}z_{i}$, where $z_{i}$ is the fractional energy of $i$-th hadron,
$n=1,2,3$ is the number of hadrons observed in the final state; (v) the dependence
on $n$, where the fractional energies of the all observed hadrons are integrated in
the region $0.1<z<0.33$.\\
%%%%%%%%%%%%%%%%%%%%%%%%  begin thr_fig2.eps  %%%%%%%%%%%%%%%%%%%%%%%%%%%%
%\begin{figure}[!htb]
\begin{figure}[!t]
\begin{center}
\epsfxsize=8.cm
\epsfbox{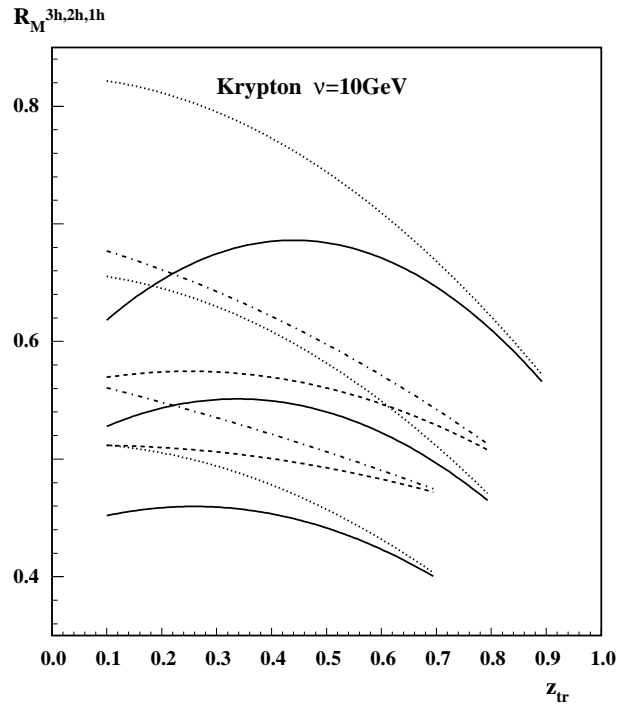}
\end{center}
\caption{\label{xx2}
{\it Ratio $R_M^{3h,2h,1h}$ for Krypton target at the energy of the virtual
photon $\nu=10GeV$ as a function of the fractional energy of the "trigger" hadron 
$z_{tr}$. Solid lines correspond to the case of full attenuation. From up to down
the ratios $R_M^{1h}$, $R_M^{2h}$ and $R_M^{3h}$ are presented, respectively.
Dashed lines correspond to the case of partial attenuation. From up to down the
ratios $R_M^{2h}$ and $R_M^{3h}$ are presented, respectively. Dotted lines
correspond to the case of full attenuation and adjacent hadrons produced on the
fast end of the string. From up to down the ratios $R_M^{1h}$, $R_M^{2h}$ and
$R_M^{3h}$ are presented, respectively. Dot-dashed lines correspond to the case of 
partial attenuation and adjacent hadrons produced on the fast end of the string.
From up to down the ratios $R_M^{2h}$ and $R_M^{3h}$ are presented, respectively.
}}
\end{figure}
%%%%%%%%%%%%%%%%%%%%%%%%  end thr_fig2.eps  %%%%%%%%%%%%%%%%%%%%%%%%%%%%
%%%%%%%%%%%%%%%%%%%%%%%%  begin thr_fig3.eps  %%%%%%%%%%%%%%%%%%%%%%%%%%%%
\begin{figure}[!htb]
\begin{center}
\epsfxsize=8.cm
\epsfbox{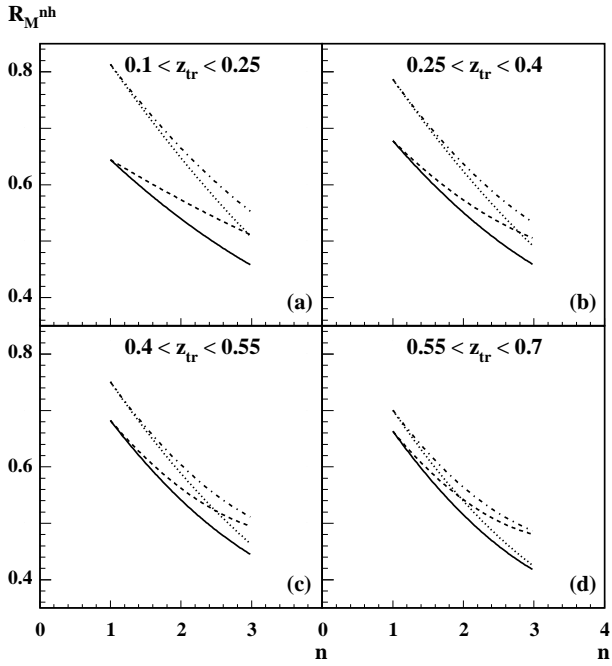}
\end{center}
\caption{\label{xx3}
{\it Ratio $R_M^{nh}$ for the Krypton target at the energy of the virtual photon
$\nu=10GeV$ as a function of $n$, where the fractional energy of the "trigger"
hadron is averaged in the region: (a) $0.1<z_{tr}<0.25$; (b) $0.25<z_{tr}<0.4$;
(c) $0.4<z_{tr}<0.55$; and (d) $0.55<z_{tr}<0.7$. The fractional energies of other
hadrons are integrated over kinematically allowed regions.
Notations are the same as Fig.2.
}}
\end{figure}
%%%%%%%%%%%%%%%%%%%%%%%%  end thr_fig3.eps  %%%%%%%%%%%%%%%%%%%%%%%%%%%%
%%%%%%%%%%%%%%%%%%%%%%%%  begin thr_fig4.eps  %%%%%%%%%%%%%%%%%%%%%%%%%%%%
\begin{figure}[!htb]
\begin{center}
\epsfxsize=8.cm
\epsfbox{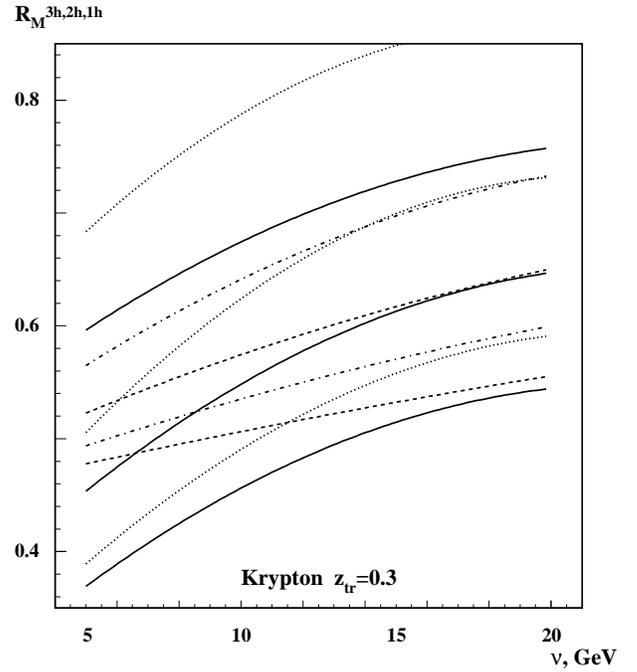}
\end{center}
\caption{\label{xx4}
{\it Ratio $R_M^{3h,2h,1h}$ for Krypton target at fractional energy of the
"trigger" hadron $z_{tr}$ fixed at $z_{tr}=0.3$ as a function of the energy of
the virtual photon $\nu$.
Notations are the same as Fig.2.
}}
\end{figure}
%%%%%%%%%%%%%%%%%%%%%%%%  end thr_fig4.eps  %%%%%%%%%%%%%%%%%%%%%%%%%%%%
%%%%%%%%%%%%%%%%%%%%%%%%  begin thr_fig5.eps  %%%%%%%%%%%%%%%%%%%%%%%%%%%%
\begin{figure}[!htb]
\begin{center}
\epsfxsize=8.cm
\epsfbox{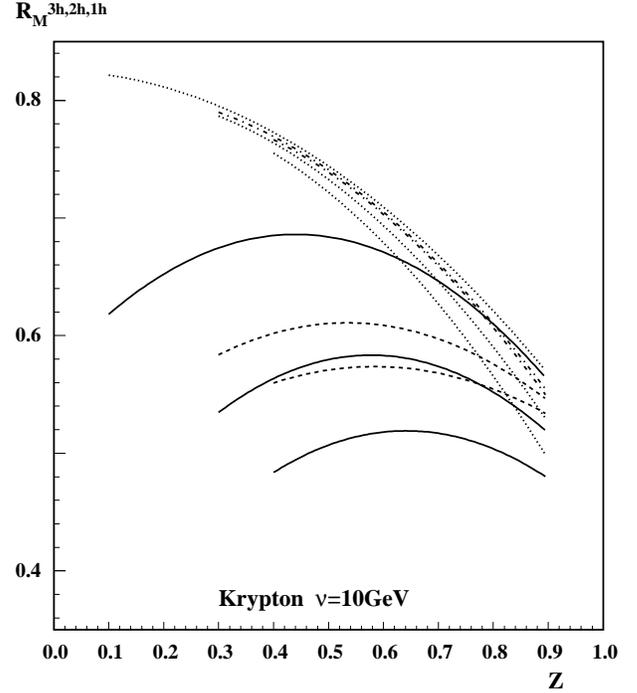}
\end{center}
\caption{\label{xx5}
{\it Ratio $R_M^{3h,2h,1h}$ for Krypton target at energy of the virtual photon
$\nu=10GeV$ as a function of $Z=\sum_{i=1}^{n}z_{i}$, where $z_{i}$ is the
fractional energy of $i$-th hadron, $n=1,2,3$ is the number of hadrons observed
in the final state.
Notations are the same as Fig.2.
}}
\end{figure}
%%%%%%%%%%%%%%%%%%%%%%%%  end thr_fig5.eps  %%%%%%%%%%%%%%%%%%%%%%%%%%%%
%%%%%%%%%%%%%%%%%%%%%%%%  begin thr_fig6.eps  %%%%%%%%%%%%%%%%%%%%%%%%%%%%
\begin{figure}[!htb]
\begin{center}
\epsfxsize=8.cm
\epsfbox{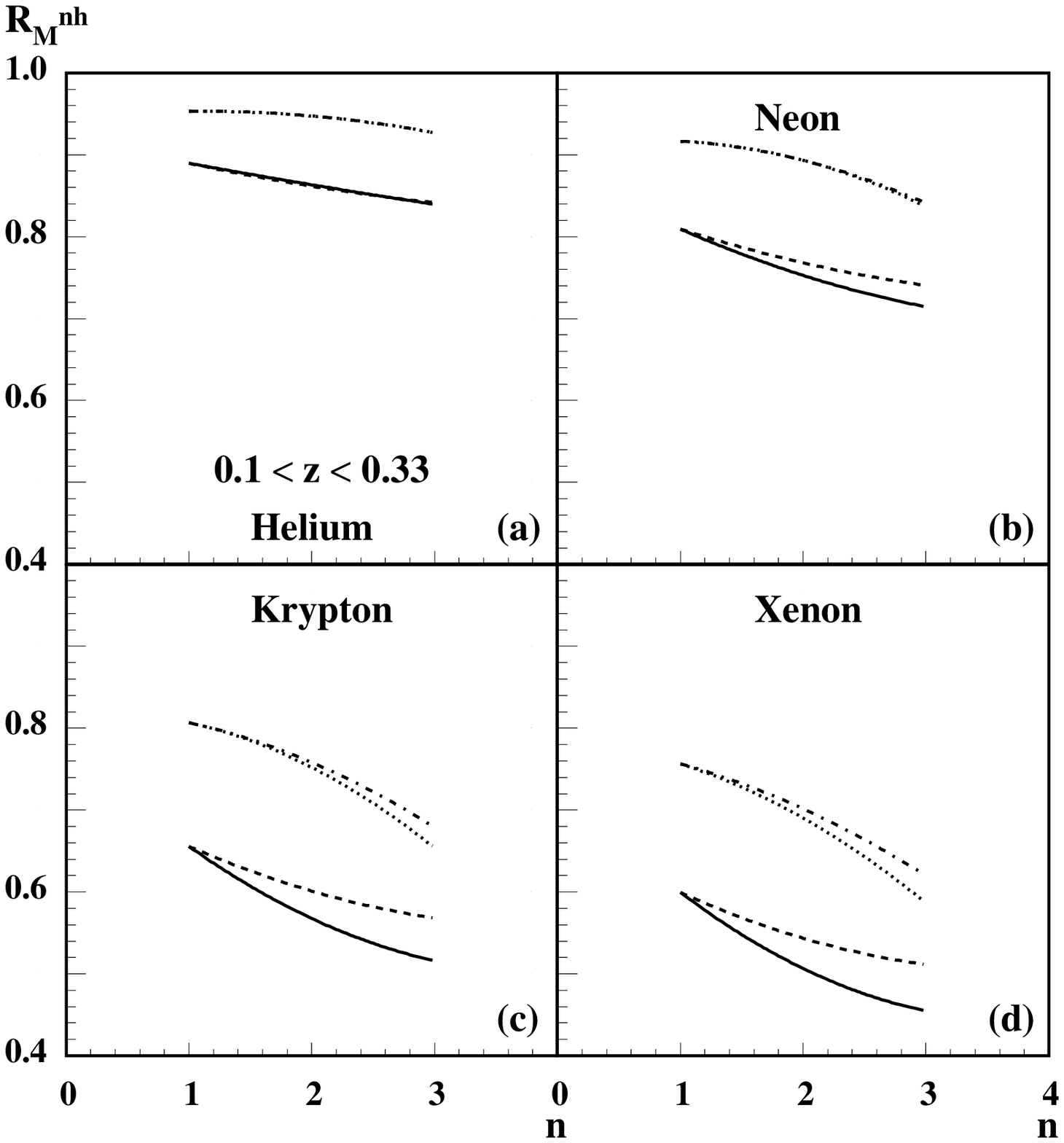}
\end{center} 
\caption{\label{xx6}
{\it Ratio $R_M^{nh}$ at energy of virtual photon $\nu=10GeV$ as a function of
$n$, where $n=1,2,3$ is the number of hadrons observed in the final state.
The fractional energies of the all hadrons are integrated in the region
$0.1<z<0.33$. The results for different nuclei are presented: (a) Helium;
(b) Neon; (c) Krypton and (d) Xenon.
Notations are the same as Fig.2.
}}
\end{figure}
%%%%%%%%%%%%%%%%%%%%%%%%  end thr_fig6.eps  %%%%%%%%%%%%%%%%%%%%%%%%%%%%
At present it is assumed that hadrons produced from one string attenuate 
independently (full attenuation). This seems strange for the following reason.
String has transverse dimensions comparable with the transverse size of the
hadrons (at least no more).
Therefore it is natural to suppose, that hadrons produced from one string, may 
partially screen one another, what in result must to lead to the weakness of NA 
(partial attenuation). For the study of this effect, and for comparison with 
the basic supposition that hadrons attenuate independently (full attenuation), 
we consider partial attenuation in extreme case, when hadrons fully screen 
one another, and in result multi-hadron system attenuates as a single hadron.
In accordance with above suppositions we consider four different cases for
nuclear attenuation:
(i) all parts of string and all produced hadrons are absorbed in nuclear
medium independently (full attenuation). The full attenuation corresponds eq.(4). 
In all figures for notation of full attenuation we use solid lines;
(ii) only initial string for the first produced hadron and
first produced hadron itself attenuate (partial attenuation).
Here the first produced hadron means the hadron first produced  on time among
observed ones.
The partial attenuation corresponds eq.(4) with the corresponding replacements
$Q_{2}=H_{2}=Q_{3}=H_{3}=0$.
In all figures for notation of partial attenuation we use dashed lines;
(iii) it is supposed that $n$ observed hadrons ($n=1,2,3$) are adjacent ones 
and also that they produced on the fast end of the string. It is assumed
additionally that this system suffers full attenuation in nuclear matter
\footnote{It is supposed that $n$ hadrons observed in the final state are
neighbors on the time of production, i.e. between them do not produced
additional hadrons. The term "hadrons produced on the fast end of the string"
means that it is a sequence of hadrons having lowest ranks in the string.}.
The case of $n$ adjacent hadrons produced on the fast end of the string corresponds
eq.(6), where only $\delta$-functions in square brackets are taken into account.
In all figures for notation of $n$ adjacent hadrons on the fast end of the string
and full attenuation we use dotted lines;
(iv) only $n$ adjacent hadrons produced on the fast end of the string are taken
into account. It is supposed also that only initial string for the first produced 
hadron and first produced hadron itself attenuate. In all figures for notation of
$n$ adjacent hadrons on the fast end of the string and partial attenuation we use 
dot-dashed lines.\\
\hspace*{1em}
Results of calculations with these conditions are shown in Figs.2-6.
The nuclear density functions and set of parameters
used in calculations were taken from our recent work~\cite{A19}.\\
%%% Fig.2
\hspace*{1em}
In Fig.2 the multiplicity ratios $R_M^{3h,2h,1h}$ for Krypton target at the
energy of the virtual photon $\nu=10GeV$ as a function of the fractional energy
of the "trigger" hadron $z_{tr}$ are presented.
Solid lines correspond to the case of random selection of hadrons from the jet
and full attenuation. From up to down the
ratios $R_M^{1h}$, $R_M^{2h}$ and $R_M^{3h}$ are presented, respectively.
Dashed lines correspond to the case of random selection of hadrons from the jet
and partial attenuation.
From up to down the ratios $R_M^{2h}$ and $R_M^{3h}$ are presented, respectively.
Dotted lines correspond to the case of adjacent hadrons
produced on the fast end of the string and full attenuation.
From up to down the ratios $R_M^{1h}$,
$R_M^{2h}$ and $R_M^{3h}$ are presented, respectively.
Dot-dashed lines correspond to the case of adjacent hadrons
produced on the fast end of the string and partial attenuation.
From up to down the ratios $R_M^{2h}$
and $R_M^{3h}$ are presented, respectively.\\
From Fig.2 it is easy to see that high values of $z_{tr}$ ($z_{tr} \ge 0.7$)
are very convenient for the studying the mutual screening of the hadrons.
For these values of $z_{tr}$ positions of the particles in the string do not play a 
essential role.
Mutual screening leads to the fact that the curves corresponding full and partial 
screening quite substantially different. The difference for the three-particle case
is more than for two-particle case.
The positions of the particles in the string become significant in the case of small
values of $z_{tr}$ ($z_{tr} \le 0.3$).
Particles produced on the fast end of the string are attenuated less than others. 
The largest difference occurs in the case of single hadron, the smallest difference
occurs in the case of three hadrons.\\
%%% Fig.3
\hspace*{1em}
In Fig.3 the ratio $R_M^{nh}$ for the Krypton target at the energy
of the virtual photon $\nu=10GeV$ as a function of $n$ are presented, where
$n=1,2,3$ is the number of hadrons observed in the final state.
The fractional energy of the "trigger" hadron is averaged in the region:  
(a) $0.1<z_{tr}<0.25$; (b) $0.25<z_{tr}<0.4$; (c) $0.4<z_{tr}<0.55$;
and (d) $0.55<z_{tr}<0.7$. The fractional energies of other hadrons are
integrated over kinematically allowed regions.
Notations are the same as Fig.2.
From Fig.3 we see that the study of small and medium $z_{tr}$ (panels $a, b, c$)
in the case of $n=1$ can provide information about whether the observed hadron 
leading (i.e., that it was produced on the fast end of the string and contains 
knocked out quark) or not. In the case of $n=2$ it is convenient to study positions 
of hadrons in string at small and medium $z_{tr}$ (panels $a, b, c$). 
In the case of $n=3$ the study of medium and large $z_{tr}$ (panels $c, d$) can
show the screening in the three-particle system takes place or not.\\
%%% Fig.4
\hspace*{1em}
In Fig.4 the ratios $R_M^{3h,2h,1h}$ for Krypton target at fractional energy of the
"trigger" hadron $z_{tr}$ fixed at $z_{tr}=0.3$ as a function of the energy of the 
virtual photon $\nu$ are presented.
Notations are the same as Fig.2.
Fig.4 shows that our questions: (i) attenuation is full or partial, (ii) observed
particles are neighbors produced on the fast end of the string or not
are, can be examined in the entire region considered energies. However, question
(i) it is convenient to consider in the region relatively low energies ($\nu \sim 5 
GeV$), question (ii) in the region relatively high energies ($\nu \sim 20 GeV$).\\
%%% Fig.5
\hspace*{1em}
In Fig.5 the ratios $R_M^{3h,2h,1h}$ for Krypton target at energy of the virtual
photon $\nu=10GeV$ as a function of $Z=\sum_{i=1}^{n}z_{i}$, where
$z_{i}$ is the fractional energy of $i$-th hadron are presented.
Notations are the same as Fig.2.
By this variable it is convenient to explore the question: are the produced
particles neighbors produced on the fast end of the string or not? Behavior of the
systems containing only the neighboring particles on the fast end of the string
is qualitatively different from behavior of the
systems that contain them among others.\\
%%% Fig.6
\hspace*{1em}
In Fig.6 the ratios $R_M^{nh}$ at energy of virtual photon $\nu=10GeV$
as a function of $n$ are presented.
The fractional energies of the all observed hadrons are integrated in the region 
$0.1<z<0.33$. The results for different nuclei are presented: (a) Helium;
(b) Neon; (c) Krypton and (d) Xenon.
Notations are the same as Fig.2.
It is easy to see that nuclear effects are amplified with the increasing of the atomic 
mass number $A$. The joint experimental study of the one-, two- and three-hadron
systems can be very helpful.
%%%
We want to note that in the energy region studied in this work (5-20GeV) the
number of hadrons in the current fragmentation region is limited and the
probability to have among the observed hadrons two 
or even three adjacent ones is large. Also, it is likely that these hadrons 
were produced on the fast end of the string.\\
\hspace*{1em}
Also we would like to briefly discuss why the cross section of the string-nucleon
interaction may be equal to the cross section of hadron-nucleon interaction.
Since the string is an object with small transverse dimensions the probability of 
mutual screening of the particles in the string is very large.\\
\hspace*{1em}
There are other reasons why multi-hadron systems can attenuate as a single hadron.
The two- or three-hadron systems will attenuate as a single hadron when
final hadrons appear in result of decay of one resonance. For instance, 
combinations two or three pions can be obtained in result of decay
of single vector meson produced in nucleus and decayed behind it.\\
%%%%%%%%%%%%%%%  end \section{Results and Discussion}  %%%%%%%%%%%

%%%%%%%%%%%%%%%  begin \section{Conclusions}  %%%%%%%%%%%
\section{Conclusions}
\hspace*{1em}
In this paper the problem of mutual screening of the prehadrons
and hadrons in string in the framework of standard Lund model has been considered.
We have shown that if the relevant data will be obtained, it will assess the degree
of mutual screening of the particles in the string.
From our point of view, such information would be very useful for understanding the 
behavior of jets in high energy hadron-nucleus and nucleus-nucleus interactions.
Unfortunately, many questions remained over the scope of this work:
(i) How strongly the results depend on the chosen model?
As mentioned above, a simple formula for scaling function in the standard Lund model
allows to obtain expressions for any number of hadrons in a compact form.
Such compact expression can not be obtained in the case of more complex 
scaling functions (for instance, symmetric Lund model's scaling function).
(ii) How the results change if we consider hadrons with specific charges and the
different cross-sections?
We can choose a combination of particles that can not be neighbors in the string, or
have a very different cross sections. Experimental study of such combinations can be
very useful for the
development of the model. (iii) How the results change if we consider that two or
three hadrons could be produced as a result of decay of the single resonance?
Above we tried qualitatively answer this question, but a quantitative study is 
needed.
(iv) In this work the basic case was considered, when in the direction of virtual
photon, in result of DIS, one string arises. In the work~\cite{A20} we considered
the case, when in the direction of virtual photon arise both one and two strings. 
Contribution of the events with two strings is relatively small in the case of
single hadron. However it can essentially increase if in the final state are observed
two or three hadrons. These questions will be discussed in further publications.
\begin{acknowledgments}
The author would like to thank H. Gulkanyan for fruitful discussions. This work has 
been partially supported by Cooperation Agreement between DESY and YerPhI.
\end{acknowledgments}
%%%%%%%%%%%%%%%  end \section{Conclusions}  %%%%%%%%%%%

\end{document}